\documentclass[final]{aipproc}

\usepackage{epsfig}

\layoutstyle{8x11single}

\def\ket#1{|#1\rangle}

\def\scal#1#2{\langle#1|#2\rangle}
\def\matr#1#2#3{\langle#1|#2|#3\rangle}

\def\ave#1#2{\langle #1\rangle_{#2}}

\begin{document}

\title{Thermodynamic Analogy for Structural Phase Transitions}


\author{Pavel Cejnar}{
  address={Institute of Particle and Nuclear Physics, Charles University,
  V Hole{\v s}ovi{\v c}k{\'a}ch 2, 180\,00 Prague, Czech Rep.}
}
\author{Stefan Heinze}{
  address={Institute of Nuclear Physics, University of Cologne,
  Z\"ulpicherstrasse 77, 50937 Cologne, Germany}
}
\author{Jan Dobe{\v s}}{
  address={Nuclear Physics Institute, Academy of Sciences of the Czech Republic, 
  250\,68 {\v R}e{\v z}, Czech Rep.}
}

\begin{abstract}
We investigate the relationship between ground-state (zero-temperature) quantum 
phase transitions in systems with variable Hamiltonian parameters and classical
(temperature-driven) phase transitions in standard thermodynamics.
An analogy is found between (i) phase-transitional distributions of the 
ground-state related branch points of quantum Hamiltonians in the complex
parameter plane and (ii) distributions of zeros of classical partition 
functions in complex temperatures.
Our approach properly describes the first- and second-order quantum 
phase transitions in the interacting boson model and can be generalized 
to finite temperatures.
\end{abstract}

\maketitle


\section{Introduction}
\label{intro}

Quantum structural phase transitions (QPT's) at zero temperature are well 
known in both lattice \cite{Sachdev} and many-body systems
\cite{Gil}--\cite{Cej5}.
In any of such transitions, the structure of the ground state (and few 
low-lying states) does not evolve smoothly with the Hamiltonian control
parameters, but flips abruptly from one configuration to another at 
a certain \lq\lq critical point\rq\rq.
A typical QPT Hamiltonian reads as 
\begin{equation}
H(\lambda)=H_0+\lambda V=(1-\lambda)H(0)+\lambda H(1)\ ,
\label{Hamg}
\end{equation}
where $H_0$ and $V$ represent two incompatible terms, $[H_0,V]\neq 0$,
and $\lambda$ is a dimensionless control parameter.
Since arbitrary scaling and sign factors can be absorbed in $V$, one 
can require $\lambda\in[0,1]$.
The variation of $\lambda$ drives the system between two limiting
$\lambda=0$ and $\lambda=1$ modes of motions, which are in the 
many-body case typically associated with spherical and deformed 
shapes of nuclei \cite{Diep,Mor,Cas,Ia1,Jol2,Cejle,Jol3,Cap}, 
paired and unpaired 
phases of strongly interacting Fermi systems \cite{Zha,Heiss1,Rowe,Vol}, 
or with normal and superradiant modes of some quantum optical systems 
\cite{Ema}.

For $|\lambda|$ large enough, the term $\lambda V$ in Eq.~(\ref{Hamg})
prevails over $H_0$ and the spectrum of $H(\lambda)$ just freely 
expands.
The most interesting physics thus happens around the minimum $\lambda_0$ 
of the parabola that for finite Hilbert-space dimensions $n$ determines 
the dispersion $\Delta^2E=n^{-1}{\rm tr}\,H^2-n^{-2}{\rm tr}^2H$ of 
energies in the spectrum as a function of $\lambda$.
The ground-state (g.s.) related structural phase transitions (if any) 
most typically appear at a critical value $\lambda_{\rm c}$ not far 
away from $\lambda_0$.
They can be observed by analyzing the average
\begin{equation}
\ave{V}{0}\equiv\matr{\Psi_0}{V}{\Psi_0}=\frac{dE_0}{d\lambda}\ ,
\end{equation}
where $E_0(\lambda)$ is the energy and $\ket{\Psi_0(\lambda)}$ the
normalized eigenvector of the ground state.
It can be shown that $\ave{V}{0}$ is a nonincreasing function of 
$\lambda$, which in the QPT case develops either a discontinuity 
or nonanalyticity at $\lambda_{\rm c}$.
The $n\to\infty$ limit is usually required for a QPT to occur 
(diagonalization of a finite-$n$ Hamiltonian cannot generate 
a nonanalytic behavior), but distinctive QPT precursors can be often 
observed already in moderate dimensions.
(Note, however, that a discontinuous change of the $\ave{V}{0}$ can 
take place even in finite-$n$ cases if the ground state undergoes 
an unavoided crossing with another state \cite{Ari}.)
If the $(\kappa-1)$th derivative of $\ave{V}{0}$ is discontinuous
at $\lambda_{\rm c}$, the derivatives of $E_0$ are discontinuous 
(singular) starting from the $\kappa$th one.
This situation is described as a QPT of order $\kappa$ \cite{Gil,Feng}, 
in analogy with the well-known Ehrenfest classification of thermodynamic 
phase transitions.

Of particular interest are the situations when $\ave{V}{0}$ drops
to zero and, simultaneously, also the dispersion, $\ave{V^2}{0}$,
vanishes at the critical point.
Typically, this may happen if $V$ is semi-positively definite. 
Consequently, $\ave{V}{0}$ must remain zero for all $\lambda\geq
\lambda_{\rm c}$ and the g.s. wave function gets fixed (in the 
non-degenerate case).
In these cases $\ave{V}{0}$ may be considered as an \lq\lq order 
parameter\rq\rq\ that distinguishes two quantum \lq\lq phases\rq\rq\ 
of the model (the values $\ave{V}{0}=0$ and $\ave{V}{0}\neq 0$ being 
attributed to \lq\lq more symmetric\rq\rq\ and \lq\lq less 
symmetric\rq\rq\ phases, respectively).

The ground-state QPT's happen at temperature $T=0$ and thus have no real 
thermal attributes.
Therefore, questions often arise whether the term \lq\lq phase 
transition\rq\rq\ used in this context is just a metaphor, or whether 
it refers to some deeper analogies in standard thermodynamics
\cite{Zel}.
Of course, the situation becomes more general if the system is described 
with both $\lambda$ and $T$ taken as free parameters.
The g.s. average $\ave{V}{0}$ is then replaced by a thermal average
\begin{equation}
\ave{V'}{T}\equiv\frac{{\rm tr}\,(\,V'\,e^{-T^{-1}H})}
{{\rm tr}\,e^{-T^{-1}H}}
=\frac{\partial F_0}{\partial\lambda}\ ,
\label{finiT}
\end{equation} 
where $F_0(\lambda,T)=-T\ln\,{\rm tr}\,e^{-T^{-1}H(\lambda)}$ is the 
equilibrium value of the free energy and $V'=-T[\frac{\partial}
{\partial\lambda}e^{-T^{-1}H(\lambda)}] e^{T^{-1}H(\lambda)}$ 
an operator generating the finite-temperature \lq\lq order 
parameter\rq\rq\ (one can prove that $\ave{V'}{T}\to\ave{V}{0}$ for 
$T\to 0$).
In this case, both structural (driven by $\lambda$) and thermodynamic 
(driven by $T$) phase transitions turn out to be just two different 
aspects of the same phenomenon, residing in the full $\lambda\times T$ 
parameter space \cite{Sachdev,Gil}, and a common language should be
developed for their description.
 
In this contribution, we will mostly deal only with the simplest case, 
i.e., with the $T=0$ limit of quantum phase transitions.
The approach \cite{Cej5} will be presented that makes it 
possible to treat such transitions fully in parallel with thermodynamic 
phase transitions at finite temperatures.
This approach is based on a similarity between the distribution of zeros 
of the partition function $Z$ at complex temperatures 
for systems undergoing classical phase transitions \cite{Yang,Gross} 
and the distribution of so-called branch points of QPT Hamiltonians 
(\ref{Hamg}) in complex-extended $\lambda$ plane \cite{Heiss1}.
Since a generalization of our method to finite temperatures seems possible, 
we believe that it represents the right way toward the unified description  
of quantum phase transitions in the $\lambda\times T$ space.

\section{QPT's in the interacting boson model}

As a testing ground, we will use the interacting boson model (IBM) 
\cite{Iach}, well known from nuclear-structure studies as well as 
from analyses focused on general features of quantum chaos and 
quantum phase transitions.
This model describes nuclear shapes and collective motions in terms of 
an ensemble of $N$ interacting $s$ and $d$ bosons with angular 
momenta 0 and 2, respectively.
Its algebraic formulation is based on the dynamical algebra U(6) 
formed by 36 bilinear products $b_i^{\dagger}b_j$ of boson creation 
and annihilation operators, where $b_i=s$ or $d_{\mu}$ with 
$\mu=-2,\dots,+2$.
This allows one to extract several alternative dynamical symmetries and
find analytic solutions for the corresponding Hamiltonians.
All the IBM dynamical symmetries lead to the algebra O(3) of angular 
momentum $L=\sqrt{10}(d^{\dagger}d)^{(1)}$ which guarantees the invariant 
symmetry of the model under rotations.

Individual dynamical symmetries are named after the first algebras in
the respective U(6)$\supset\dots\supset$O(3) decompositions, so we have
U(5), SU(3), O(6), $\overline{\rm SU(3)}$, and $\overline{\rm O(6)}$
dynamical symmetries (the fourth and fifth chain differs from the
second and third one, respectively, just by relative phases between
$s$ and $d$ bosons in definitions of the corresponding algebras).
Both first- and second-order QPT's---according to the Ehrenfest
classification---are present in the IBM parameter space between these 
symmetries.
Since geometry can be attributed to given algebraic structures 
via the coherent state formalism, the IBM phase transitions describe 
changes of the nuclear ground-state shapes.
In particular, using the projected coherent states \cite{Gin} 
\begin{equation}
\ket{N,\beta,\gamma}\propto\left(
s^{\dagger}+\beta\cos\gamma\ d^{\dagger}_0+
\frac{\beta\sin\gamma}{\sqrt{2}}[d^{\dagger}_{-2}+d^{\dagger}_{+2}]
\right)^N\ket{0}
\end{equation}
with $\beta$ and $\gamma$ interpreted as Bohr 
geometric variables, one obtains the following shape types associated 
with individual dynamical symmetries: spherical [U(5)], prolate 
[SU(3)], oblate [$\overline{\rm SU(3)}$], and deformed $\gamma$-soft 
[O(6), $\overline{\rm O(6)}$].

The IBM Hamiltonian is assumed to have a general form containing 
one- and two-body terms, with few restrictions resulting from
fundamental symmetry requirements.
This leads to the most general Hamiltonian with six free parameters 
(except an additive constant).
However, a more involved analysis focused on phase-transitional 
properties \cite{Mor} reveals that there are only two essential 
parameters.
An archetypal two-parameter Hamiltonian (see, e.g., 
Refs.~\cite{Cej1,Jol1}) reads as
\begin{equation}
H_{\chi}(\lambda)=(1-\lambda)\left[-\frac{Q_{\chi}\cdot Q_{\chi}}
{N}\right]+\lambda\,n_{\rm d}\ ,
\label{ham}
\end{equation}
where the parameters change within domains $\lambda\in[0,1]$ and
$\chi\in[\frac{-\sqrt{7}}{2},\frac{+\sqrt{7}}{2}]$, while 
$n_{\rm d}=d^{\dagger}\cdot{\tilde d}$ represents the $d$-boson 
number operator and $Q_{\chi}=d^{\dagger}{\tilde s}+s^{\dagger}
{\tilde d}+\chi(d^{\dagger}{\tilde d})^{(2)}$ the quadrupole 
operator.
The U(5) dynamical symmetry is located at $\eta=1$ and $\chi$ 
arbitrary, O(6) at $(\eta,\chi)=(0,0)$, SU(3) and $\overline{\rm SU(3)}$ 
at $(\eta,\chi)=(0,-\frac{\sqrt{7}}{2})$ and $(0,+\frac{\sqrt{7}}{2})$, 
respectively, while $\overline{\rm O(6)}$ is absent in the given
parametrization.

Clearly, if $\chi$ is fixed to a constant, the Hamiltonian in 
Eq.~(\ref{ham}) has the general form (\ref{Hamg}) with 
\begin{equation}
V=n_{\rm d}+\frac{Q_{\chi}\cdot Q_{\chi}}{N}
\end{equation} 
being a semi-positive operator.
It is not difficult to see that $\langle V\rangle_0>0$ for $\lambda=0$
and $\langle V\rangle_0=0$ for $\lambda=1$, so the scenario described
in the previous section may be expected to apply in the $N\to\infty$ 
limit.
Indeed, the order parameter can be expressed in terms of the Bohr 
deformation parameters $\beta_0$ and $\gamma_0$, resulting from 
minimization of the coherent-state energy functional 
\begin{equation}
{\cal E}(\lambda,\chi;\beta,\gamma)\equiv\lim_{N\to\infty}
\frac{1}{N}\,\ave{H_{\chi}(\lambda)}{\ket{N,\beta,\gamma}}=
\frac{(5\lambda-4)\beta^2+4\sqrt{\frac{2}{7}}(1-\lambda)\beta^3
\cos 3\gamma+[\lambda-\frac{2}{7}\chi^2(1-\lambda)]\beta^4}
{(1+\beta^2)^2}
\label{Efun}
\end{equation} 
(normalization per boson is included to ensure finite $N\to\infty$ asymptotics), 
namely  
\begin{equation}
\ave{{\cal V}}{0}\equiv\lim_{N\to\infty}\frac{1}{N}\,\langle V\rangle_0=
\frac{5\beta_0^2-4\sqrt{\frac{2}{7}}\chi\beta_0^3\cos3\gamma_0+
\left(\frac{2}{7}\chi^2+1\right)\beta_0^4}{(1+\beta_0^2)^2}.
\label{order}
\end{equation} 
For $\chi\neq 0$, the value of $\beta_0$ changes from $\beta_0\neq 0$ 
to 0 at
\begin{equation}
\lambda_{\rm c}=\frac{4+\frac{2}{7}\chi^2}{5+\frac{2}{7}\chi^2}\ ,
\label{lac}
\end{equation}
indicating a first-order deformed-to-spherical QPT (the order parameter
$\ave{{\cal V}}{0}$ drops from a positive value to zero). 
For $\chi=0$, the value $\beta_0\propto\sqrt{\lambda_{\rm c}-\lambda}$
valid in the left vicinity of $\lambda_{\rm c}$ continuously joins 
with $\beta_0=0$ valid above $\lambda_{\rm c}$; the corresponding QPT 
is of the second order, with the critical exponent for $\ave{{\cal V}}{0}$ 
equal to 1.
Because of the proper weighting of both terms in Hamiltonian (\ref{ham})
the finite-$N$ precursors of the QPT behavior are always located in the 
region around (\ref{lac}), independently of $N$.

Besides the spherical-deformed phase separatrix, there exists also
a separatrix at $\chi=0$ and $\lambda<\frac{4}{5}$ which corresponds
to the first-order QPT between prolate and oblate shapes, where either 
$\beta_0$ changes the sign or, equivalently, $\gamma_0$ jumps from 0 to 
$\frac{\pi}{6}$.
In this case, however, the form (\ref{Hamg}) [with $\lambda$ replaced
by $\chi$, and $H_0$ and $V$ by the corresponding expressions following
from Eq.~(\ref{ham})] can be used only locally, close to $\chi=0$.  
We will not be dealing with the prolate-oblate transition here.

An obvious way to classify the IBM deformed-spherical QPT (driven by the
interaction parameter $\lambda$) relies on the analogy between the 
$N\to\infty$ g.s. energy per boson, ${\cal E}_0(\lambda,\chi)\equiv
{\cal E}(\lambda,\chi;\beta_0,\gamma_0)$, as a function of $\lambda$, and 
the equilibrium value of the free energy, $F_0(T)$, as a function of 
temperature $T$. This leads to the Ehrenfest classification.
Using the standard definition $C=-T\frac{\partial^2}{\partial T^2}F_0$, one 
can even introduce a QPT analog of the \lq\lq specific heat\rq\rq
\begin{equation}
{\cal C}=-\lambda\frac{\partial^2{\cal E}_0}{\partial\lambda^2}=
-\lambda\frac{\partial\ave{{\cal V}}{0}}{\partial\lambda}=\lim_{N\to\infty}
\frac{2\lambda}{N}\sum_{i>0}\frac{|\ave{V}{0i}|^2}
{E_i-E_0}
\label{heat1}
\end{equation}
and show that it behaves exactly as expected for a thermodynamic phase 
transition of the respective order \cite{Cej4}.
Here and in the following, $E_i\equiv E_i(\lambda,\chi)$ stands
for the $i$th eigenvalue of $H_{\chi}(\lambda)$ while 
$\ave{V}{ij}\equiv\matr{\Psi_i(\lambda,\chi)}{V}{\Psi_j(\lambda,\chi)}$ 
and $\ave{V}{i}\equiv\ave{V}{ii}$ denote matrix elements of $V$ 
involving $i$th and $j$th eigenstates.
Note that the sum in Eq.~(\ref{heat1}) runs effectively only over those
states that have the same symmetry quantum numbers as the ground state,
since otherwise $\ave{V}{0i}=0$.

\section{Alternative definitions of a QPT \lq\lq specific heat\rq\rq}

It turns out that Eq.~(\ref{heat1}) does not represent the only form 
of the QPT analog of specific heat.
Inspired by the thermodynamic relation $C=T\frac{\partial}{\partial T}S$, 
where $S$ is the entropy, one can define 
${\cal C}=\lambda\frac{\partial}{\partial\lambda}{\cal S}$ with
${\cal S}=-\sum_i|\scal{i}{\Psi_0}|^2\ln|\scal{i}{\Psi_0}|^2$ being 
the wave-function entropy of the ground state with respect to the 
eigenbasis $\ket{i}$ of $H(1)$, i.e., the basis associated with the 
U(5) dynamical symmetry in our case.
This seems to be a plausible alternative definition of specific heat 
\cite{Cej4} for systems where $\ave{V}{0}$ and $\ave{V^2}{0}$ 
drop to zero at $\lambda_{\rm c}$. (Note that $\lambda$ should be 
inverted here to $\tilde{\lambda}=1-\lambda$ in order that the entropy 
${\cal S}$ increases with \lq\lq temperature\rq\rq\ $\tilde{\lambda}$.)

\begin{figure}
\epsfig{file=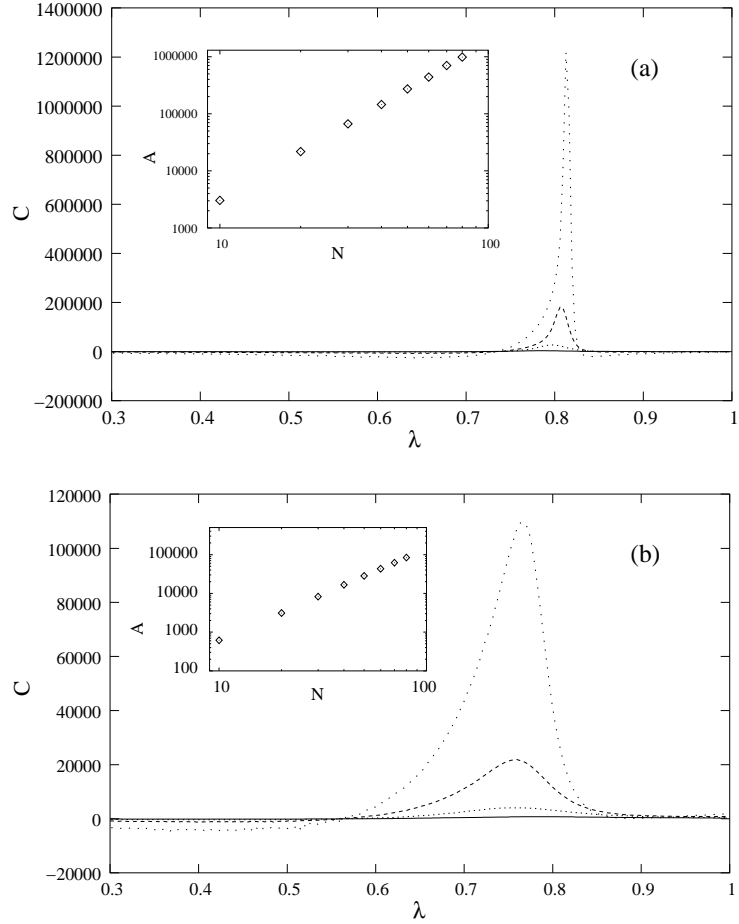,width=10cm}
\caption{\protect \lq\lq Specific heat\rq\rq\ (\ref{heat}) 
with $\Omega=1$ including all $J=0$ states for (a) the first-order 
and (b) second-order QPT of the interacting boson model [Hamiltonian 
(\ref{ham}) with (a) $\chi=\pm\frac{\sqrt{7}}{2}$ and (b) $\chi=0$]. 
The curves in order from the lowest to the highest correspond to 
$N=10$, 20, 40, and 80. The insets show the increase of the peak
maximal amplitude with $N$.} 
\label{figheat}
\end{figure}

Another possibility is to randomize Hamiltonian (\ref{Hamg}) by
adding a small stochastic component $\delta\lambda$ to the control 
parameter (with $\langle\delta\lambda\rangle=0$ and
$\langle\delta\lambda^2\rangle\equiv\sigma^2\ll 1$) and to exploit the 
perturbative expansion $H(\lambda+\delta\lambda)=H(\lambda)+
(\delta\lambda)V$.
In this way, the ground state wave-function transforms into a density 
operator $\varrho_0(\lambda)$, whose identification with a canonical 
ensemble provides a natural basis for a noise-induced thermalization 
of the given quantum system.
The resulting specific heat reads as  
\begin{equation}
{\cal C}={\rm tr}(\varrho_0\ln^2\varrho_0)-{\rm tr}^2(\varrho_0\ln\varrho_0)
\approx(\sigma^2\ln^2\sigma^2)\sum_{i>0}\frac{|\ave{V}{0i}|^2}{(E_i-E_0)^2}
\ ,
\label{rand}
\end{equation}
where the last equality is valid for very small values of $\sigma^2$
(in the IBM, a natural scaling of $\sigma^2$ for $N\to\infty$ is
$\sigma^2\propto N^{-2}$ \cite{Cej4}).
Again, formula (\ref{rand}) evaluated for the IBM Hamiltonian (\ref{ham}) 
is peaked just around $\lambda_{\rm c}$, which indicates an increased 
mixing of the Hamiltonian eigenfunctions in the QPT region.
See Refs.~\cite{Cej2,Cej4} for details.

The last analog of specific heat will be elaborated in the rest of this 
contribution.
It originates in the expression 
\begin{equation}
{\cal U}=-\sum_{i>0}\ln|E_i-E_0|
\label{poten}
\end{equation}
for the \lq\lq potential energy\rq\rq\ of a given set of levels with
the same symmetry quantum numbers, as derived from the relation 
between one-dimensional distributions of charges in a planar 
universe and analogous distributions of eigenvalues in Gaussian 
matrix ensembles \cite{Dys}.
This relation, referred to as the static Coulomb-gas analogy, must be 
distinguished from the dynamical Coulomb-gas analogy, introduced by 
Pechukas and Yukawa \cite{Pec}, that describes eigenvalue dynamics for 
Hamiltonians of the form (\ref{Hamg}).

A quantity proportional to ${\cal U}$ in Eq.~(\ref{poten}) can be 
considered as another QPT analog of the thermodynamic potential.
In particular, if $F_0$ is associated with ${\cal F}_0=\Omega^{-1}
\lambda{\cal U}$, where $\Omega$ is a scaling constant that is to be 
discussed later, the specific heat takes the form
\begin{eqnarray}
{\cal C}=-\frac{\lambda}{\Omega}\frac{\partial^2(\lambda{\cal U})}
{\partial\lambda^2}=\frac{2\lambda}{\Omega}\sum_{i>0}\Biggl[
\frac{\lambda}{2}\Biggl\{\frac{\frac{\partial^2E_i}{\partial\lambda^2}
-\frac{\partial^2E_0}{\partial\lambda^2}}{E_i-E_0}-\left(\frac{\frac{
\partial E_i}{\partial\lambda}-\frac{\partial E_0}{\partial\lambda}}
{E_i-E_0}\right)^2\Biggr\}+\frac{\frac{dE_i}{d\lambda}-\frac{\partial E_0}
{\partial\lambda}}{E_i-E_0}\Biggr]
=\qquad\qquad\qquad
\label{heat}\\
\frac{2\lambda^2}{\Omega}\left[
\sum_{i>0}\sum_{j\neq i}\frac{|\ave{V}{ij}|^2}
{(E_i-E_j)(E_i-E_0)}-
\sum_{i>0}\sum_{j>0}\frac{|\ave{V}{0j}|^2}
{(E_j-E_0)(E_i-E_0)}-
\frac{1}{2}\sum_{i>0}\left(\frac{\langle V\rangle_i-\langle V\rangle_0}
{E_i-E_0}\right)^2+
\frac{1}{\lambda}\sum_{i>0}\frac{\langle V\rangle_i-\langle V\rangle_0}
{E_i-E_0}\right]\ ,
\nonumber
\end{eqnarray}
with the last expression resulting from the Pechukas-Yukawa equations
[the last two expressions in Eq.~(\ref{heat}) are given here just for 
comparison with Eqs.~(\ref{heat1}) and (\ref{rand}); from the 
computational viewpoint the most convenient definition is of course 
represented by the first formula].
Since the motivation for the proposed relation between ${\cal F}_0$ and 
${\cal U}$ may be unclear at this moment, we ask the reader for patience 
till the next section, where the foundation of Eq.~(\ref{heat}) will 
become clear.

It turns out \cite{Cej5} that the specific heat given by the last formula 
exhibits a form that is very similar to that of Eq.~(\ref{heat1}).
This is illustrated in Figure~\ref{figheat}, where specific heat 
(\ref{heat}) with $\Omega=1$ is shown for the IBM first- and 
second-order phase transitions.
Since the $N\to\infty$ limit is numerically inaccessible, we present 
here only calculations for $N=10$, 20, 40, and 80, corresponding
to the curves with increasing height, the maximal values of ${\cal C}$ 
being shown separately in the log-log insets as a function of $N$.
Clearly, the peaks in panel (a) are sharper and higher than those
in panel (b), as indeed expected in the first- and second-order phase 
transition.
Moreover, as will be argued below, the inclusion of the right
dependence of the scaling $\Omega$ on dimension $n$ leads to the 
correct $N\to\infty$ asymptotics of Eq.~(\ref{heat}).
In this case, the specific-heat values at $\lambda_{\rm c}$ converge 
or diverge, respectively, for the IBM second- or first-order QPT's.

All the above-proposed definitions of specific heat lead to certain
sums of squared interaction matrix elements divided by some 
combinations of energy differences.
The observed peaked behavior of these quantities results from the fact 
that the spectrum becomes more compressed and, simultaneously, also the 
sum of level interactions get stronger in the phase-transitional region.
In fact, we are dealing here with a phenomenon of multiple avoided 
crossing of levels \cite{Cej1}.
The analytic description of such effects can be given in terms of 
so-called branch points of Hamiltonian (\ref{Hamg}) in the complex 
$\lambda$ plane \cite{Kato}--\cite{Rot2}.
As will be shown below, this description is particularly relevant
for the last definition of specific heat in Eq.~(\ref{heat}).

\section{Branch points and zeros of partition function}

Branch, or exceptional points are places in the complex plane of parameter 
$\lambda$ where various pairs of eigenvalues of the complex-extended 
Hamiltonian (\ref{Hamg}) coalesce \cite{Kato}.
They are simultaneous solutions of equations $\det[E-H(\lambda)]=0$ and 
$\frac{\partial}{\partial E}\det[E-H(\lambda)]=0$, that after elimination
yield the condition \cite{Zirn,Heiss2}
\begin{equation}
{\cal D}=\prod_k{\cal D}_k= 
(-)^{\frac{n(n-1)}{2}}\prod_{i<j}(E_j-E_i)^2=0
\label{disc}
\end{equation}
with ${\cal D}_k=\prod_{i(\neq k)}(E_i-E_k)$.
The discriminant ${\cal D}$ is a polynomial of order $n(n-1)$ in 
$\lambda$ (the dimension of the Hilbert space $n$ being now assumed 
finite) with real coefficients and its roots thus occur as 
$\frac{n(n-1)}{2}$ complex conjugate pairs.
Except at these points, the complex eigenvalue $E(\lambda)$ obtained 
from the characteristic polynomial of Hamiltonian (\ref{Hamg}) is 
a single analytic function defined on $n$ Riemann sheets.
The energy labels in Eq.~(\ref{disc}) enumerate the
respective Riemann sheet according to the ordering of energies at 
real $\lambda$.
The degeneracy points are square-root branch points where the Riemann 
sheets are pairwise (in generic cases) connected. 
The leading-order behavior on the two connected 
sheets close to the branch point $\lambda_{\rm b}$ is given by 
$E(\lambda)-E(\lambda_{\rm b})\approx a\sqrt{\lambda-\lambda_0}$ 
(as a doubly-valued function), with $a$ being a complex constant 
\cite{Heiss1,Shan,Heiss2}.

The relation of branch points to QPT's has been declared several 
times---see, e.g., Refs.~\cite{Heiss1,Cej1}.
Clearly, a branch point located close to the real $\lambda$ axis 
affects the local evolution of the corresponding pair of real energies 
so that the two levels undergo an avoided crossing with accompanying 
rapid changes of wave functions. 
A cumulation of branch points close to some real point $\lambda_{\rm c}$ 
thus can give rise to massive structural changes of eigenstates, as 
observed in QPT's.
It moreover turns out that the density of branch points close to 
$\lambda_{\rm c}$ may be determinative for the QPT order, as observed
on the real $\lambda$ axis.

\begin{figure}
\epsfig{file=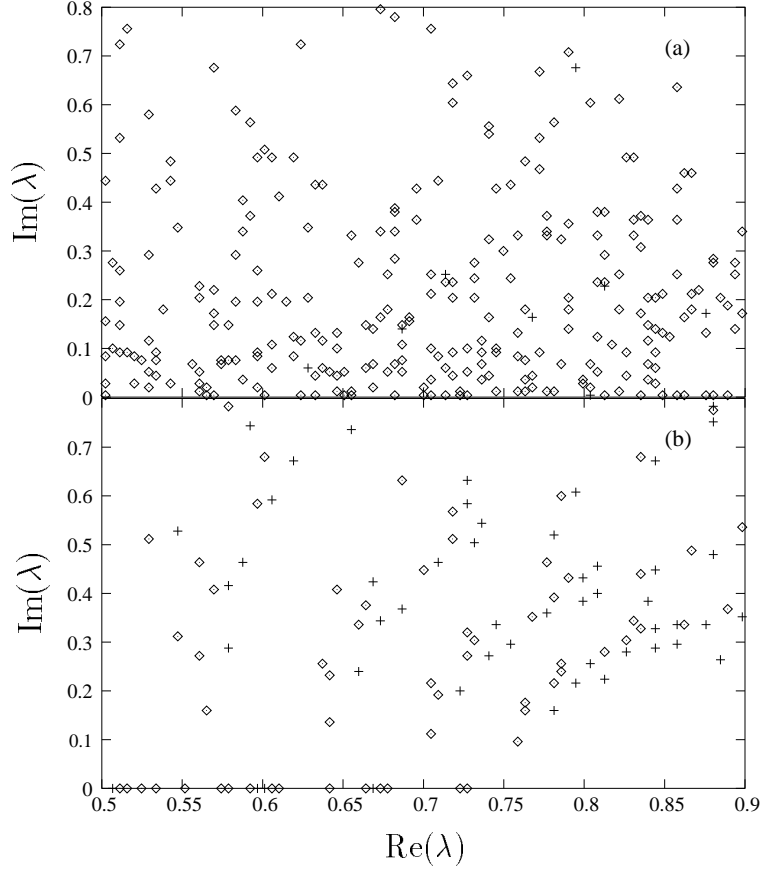,width=10cm}
\caption{\protect Branch points corresponding to all $J=0$ 
states for (a) the first-order ($|\chi|=\frac{\sqrt{7}}{2}$) and 
(b) second-order ($\chi=0$) QPT paths of the IBM Hamiltonian 
(\ref{ham}) with $N=20$. Diamonds represent single branch points
and crosses the degenerate ones (within given numerical precision).} 
\label{figex}
\end{figure}

The simplest example can be found in dimension $n=2$.
A general real 2$\times$2 Hamiltonian of the form (\ref{Hamg}) reads as
\begin{equation}
H(\lambda)=\left(
\begin{array}{cc}
a_1+a_2\lambda & c_1+c_2\lambda \\
c_1+c_2\lambda & b_1+b_2\lambda
\end{array}
\right)\ .
\label{ham2}
\end{equation}
It is easy to show that the spacing between both eigenvalues 
$E_1-E_2\equiv\Delta E$ behaves according to $|\Delta E|=
\sqrt{A+B\lambda+C\lambda^2}$, where the coefficients $A\geq 0$, 
$B$, and $C\geq 0$ are some combinations of constants in 
Eq.~(\ref{ham2}) such that the value $D=B^2-4AC$ is negative.
The minimal distance of both levels, $|\Delta E|=\Delta_0=
\sqrt{-\frac{D}{4C}}$, is reached at $\lambda=\lambda_0=-\frac{B}{2C}$.
If $\lambda$ is extended to the complex plane, one finds that both 
complex eigenvalues $E_1$ and $E_2$ cross in two complex-conjugate branch 
points $\lambda_{\rm b\pm}=\lambda_0\pm i\,\Delta_0$.
For $\Delta_0$ converging to zero, the branch points come to the real
axis and the avoided crossing turns into a real crossing,
implying $|\Delta E|=\sqrt{C}(\lambda-\lambda_{\rm b})$.
The line that passes both points $\lambda_{\rm b+}$ and $\lambda_{\rm b-}$ 
in the direction perpendicular to the real axis represents the locus of 
points where either real or imaginary part of $E_1-E_2$ vanishes. 
In particular, one has ${\rm Re}\Delta E=0$ for $|{\rm Im}\lambda|\geq
|{\rm Im}\lambda_{\rm b\pm}|$ and ${\rm Im}\Delta E=0$ for $|{\rm Im}\lambda|
\leq|{\rm Im}\lambda_{\rm b\pm}|$.
Only at $\lambda=\lambda_{\rm b\pm}$ both loci overlap
so that at these points ${\rm Re}\Delta E={\rm Im}\Delta E=0$.
This corresponds to the above-mentioned behavior of the complex square
root (valid for $\Delta_0>0$).

In the $n=2$ case, the domain of $E(\lambda)$ consists of two Riemann 
sheets only.
It is evident that the analysis becomes much more complicated for higher 
dimensions.
Methods to locate individual branch points in the complex plane for
arbitrary $n$ were developed (see, e.g., Ref.~\cite{Zirn}), but 
without the knowledge of explicit Riemann sheet structure of the
specific problem one cannot select those branch points that are
located on the ground-state sheet.
An example is shown in Figure~\ref{figex}.
Here, branch points corresponding to the $J=0$ submatrix of the IBM 
Hamiltonian (\ref{ham}) with $N=20$ are shown for the first-order
QPT with $\chi=\pm\frac{\sqrt{7}}{2}$ (panel a) and for second-order
transition, $\chi=0$ (panel b).
The distributions in both panels show no cumulation of branch points
near the respective critical points $\lambda_{\rm c}=0.8181\dots$ (a) 
and $\lambda_{\rm c}=0.8$ (b), which reflects the fact that all Riemann
sheets are mixed together.

As can be observed,
some of the branch points in panel (a) of Fig.~\ref{figex} come very 
close to the real axis, but the absence of additional integrals of 
motions in the $\chi\neq 0$ transitional regimes guarantees that 
${\rm Im}\lambda_{\rm b}\neq 0$ in all these cases.
On the other hand, several branch points in panel (b) are located 
exactly on the real axis, with ${\rm Im}\lambda_{\rm b}=0$, which 
seems to contradict the familiar no-crossing rule.
Indeed, the $\chi=0$ transitional path is not quite generic since 
the underlying O(5) dynamical symmetry of the O(6) and U(5) limits
remains unbroken and gives rise to the seniority quantum number $v$,
that is valid along the whole transition \cite{Levi}.
Consequently, levels with different $v$'s can cross without repulsion.
It must be stressed, however, that none of these very close avoided 
(panel a) or unavoided (panel b) crossings is related to the ground-state 
Riemann sheet.

In generic cases, the determination of the Riemann sheet structure in
its full complexity is prohibitively difficult even for moderate
dimensions.
The loci of ${\rm Re}\Delta E=0$ and ${\rm Im}\Delta E=0$ crossings
are not any more straight lines, as for $n=2$, but generate complicated
patterns that can be hardly disentangled having only a finite 
numerical precision.
(Nevertheless, the ${\rm Re}\Delta E=0$ curves for IBM form a flow
with prevailing perpendicular orientation toward the real $\lambda$
axis, see the preprint in Ref.~\cite{Cej5}.)
As a consequence, practically nothing is known about the density of 
the ground-state branch points in a vicinity of $\lambda_{\rm c}$ 
for QPT's of various orders.

We will show that the specific heat in Eq.~(\ref{heat}), although
depending solely on the real-$\lambda$ observables, represents
an {\em indirect measure\/} of the density of branch points 
on the ground-state Riemann sheet near the real axis.
To prove this, we start from following question: If ${\cal F}_0=
\Omega^{-1}\lambda{\cal U}$ represents the equilibrium value of the 
\lq\lq free energy\rq\rq, as assumed above, what 
is the corresponding \lq\lq partition function\rq\rq\ ${\cal Z}$? 
Using the thermodynamic relation $F_0=-T\ln Z$, one finds
\begin{equation}
{\cal Z}^{\Omega}=\prod_{i>0}(E_i-E_0)\ ,
\label{zzz}
\end{equation}
thus ${\cal Z}$ coincides with the $\Omega^{-1}$th power of the partial 
discriminant ${\cal D}_0$ from Eq.~(\ref{disc}).
Recall that the square ${\cal D}_k^2$ is a polynomial in $\lambda$ 
with $n-1$ complex conjugate pairs of roots, each of them being 
simultaneously assigned to one other ${\cal D}_{k'}^2$.
These roots correspond to the branch points located on the $k$-th
Riemann sheet. 
Thus branch points on the ground-state Riemann sheet are zeros of 
the fictitious partition function ${\cal Z}$, that generates the 
specific heat in Eq.~(\ref{heat}). 

Zeros of canonical or grand canonical partition functions $Z$ 
in complex temperatures and/or chemical potentials play an essential 
role in the fundamental theory of thermodynamic phase transitions.
Indeed, as noticed by Yang and Lee \cite{Yang}, for a system of particles 
with mass $m$ interacting through a potential with a hard repulsive core 
the grand partition function $Z$ can be written as a polynomial in the 
parameter $y=(\hbar^{-1}\sqrt{mT})^3\exp{(T^{-1}\mu)}$, containing both 
temperature $T$ and chemical potential $\mu$. 
While for a finite size of the system the roots of $Z(y)$ must keep away 
from the real $y$ axis, in the thermodynamic limit there may exist places 
where the zeros approach infinitely close to the real axis, giving rise 
to a phase transition at the given value $y_{\rm c}$. 
This method was adapted also for canonical ensembles \cite{Gross}, where
it was shown that the density of zeros of the partition function close to 
the phase-transitional temperature $T_{\rm c}$ even determines the order 
of the transition, as reflected by the behavior of standard thermodynamic 
specific heat.

This was recently proposed \cite{Bor} as a basis for phase-transitional 
analyses in small systems.
For instance, if zeros of $Z$ are located along a curve crossing the 
real axis and if the closest zero converges to a real point $T_{\rm c}$  
in the thermodynamic limit, the order of the corresponding phase 
transition at $T_{\rm c}$ is determined by (i) the power $\alpha$ in 
the dependence $\rho\propto({\rm Im}T)^{\alpha}$ of the density of zeros 
close to the real axis, and by (ii) the angle $\nu$ between the $Z=0$ 
domain and the normal to real axis \cite{Bor}.
Namely, the phase transition will be of the first order for 
$\alpha=\nu=0$, of the second order for $\alpha\in(0,1]$, and of
a higher order for $\alpha>1$.
A detailed analysis of even more possibilities can be found in 
Ref.~\cite{Gross}.

Because the \lq\lq specific heat\rq\rq\ (\ref{heat}) results---as we know 
now---from the \lq\lq partition function\rq\rq\ (\ref{zzz}), it 
basically measures the density of zeros of ${\cal Z}$ near the real 
$\lambda$ axis, or, in other words, the density of branch points on the 
ground-state Riemann sheet. 
The peaked behaviors shown in Fig.~\ref{figheat} thus indicate that the 
IBM g.s. branch points indeed accumulate near $\lambda_{\rm c}$ for
$N\to\infty$, and that the degree of this accumulation differs for 
transitions of different orders.
We are now in a position to formulate the central statement of this
contribution, namely, the surmise \cite{Cej5} that {\em the $\kappa$th-order 
QPT distribution of the ground-state branch points is quantitatively similar
to a distribution of the $Z(T)$ complex zeros in a thermodynamic phase 
transition of the same order $\kappa$.}

\section{Normalization and large-N asymptotics}

To test the conjecture proposed at the end of the last section, we need
to discuss the factor $\Omega$, that appears in Eqs.~(\ref{heat}) 
and (\ref{zzz}).
In fact, this factor should not be a constant, but a certain function 
depending on the system's size (quite similarly to the standard
specific heat, which must be normalized to a unit part of the 
thermodynamic system).
Since the polynomial ${\cal D}_0^2$ is determined (up to a multiplicative
constant) by its complex roots, one may express specific heat (\ref{heat})
as an integral containing the density $\rho_0(\lambda)=\sum_i\delta(\lambda-
\lambda_{{\rm b}i})$ of the branch points $\lambda_{{\rm b}i}$ on the
g.s. Riemann sheet \cite{Cej5}.
The normalization of $\int\rho_0d\lambda$ to unity yields 
a factor $\propto(n-1)^{-1}$ (there are $n-1$ branch points on each 
Riemann sheet) and one arrives at the formula
\begin{equation}
\Omega=\Omega_0(n-1)\ ,
\label{norma}
\end{equation}
where $n$ is the relevant Hilbert space dimension and $\Omega_0$ an 
arbitrary constant.
Eq.~(\ref{norma}) is analogous to the scaling based on the grand partition 
function, as discussed by Yang and Lee \cite{Yang}, since in that case the 
order of the polynomial $Z(y)$ equals to $N_{\rm max}$, the maximal number 
of particles in a given volume, and $\Omega\propto N_{\rm max}$ can be 
identified with the volume.

\begin{figure}
\epsfig{file=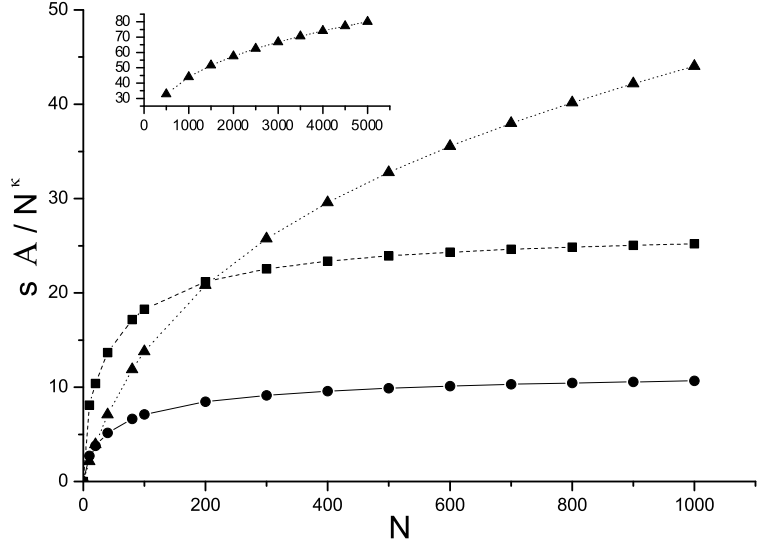,width=10cm}
\caption{\protect Maximal values of specific heat (\ref{heat}) 
(squares and triangles) normalized according to Eq.~(\ref{norma})
in the IBM second-order QPT ($\chi=0$) for very large boson numbers.
Squares (or triangles) correspond to the inclusion of the whole 
$J=0$ space (or its $v=0$ subspace) into the sum in Eq.~(\ref{heat}).
Different (arbitrary) scaling factors were used to show both sets of 
data within the same range.
For comparison, the convergence of the maximal value of specific
heat (\ref{heat1}) to its asymptotic limit $A_{\infty}=12.5$
is shown by full circles.}
\label{asymp}
\end{figure}

In the IBM, the total number of states grows roughly as $\sim N^5/120$ 
for very large boson numbers (if not counting the rotational degeneracy).
However, the dimension of the $J=0$ subspace, which is relevant in the 
calculation leading to Fig.~\ref{figheat}, is given by $n\sim N^2/12$.
So the correct normalization of Eq.~(\ref{heat}) in the IBM case is by
a factor $\Omega\propto N^2$.
Figure~\ref{asymp} shows---by the curve demarcated by squares---the maximal 
values of such normalized specific heat in the second-order ($\chi=0$) QPT 
region for very high boson numbers $N$. 
Note that such high-dimensional calculation were enabled by the underlying 
O(5) dynamical symmetry at $\chi=0$, so they could not be performed for the 
other transitional path (a).
The algebraic increase shown for $N\leq 80$ in the inset of 
Fig.~\ref{figheat}(b) (calculated for $\Omega=1$) represents only the 
initial part of the curve in Fig.~\ref{asymp}, where the convergence to 
the $A\propto N^2$ asymptotics is evident for $N>300$.
For comparison, the convergence of the maximal value of specific heat 
(\ref{heat1}) to its asymptotic value (which is for $\chi=0$ equal to
12.5) is also shown in Fig.~\ref{asymp} by full circles.
Clearly, the degree of convergence in both cases (squares and circles) 
is about the same.
On the other hand, since the $N\leq 80$ increase observed in panel (a) 
of Fig.~\ref{figheat} is much faster than the increase in panel (b), one 
expects that the asymptotic increase of the maximal ${\cal C}$ value in 
the first-order QPT is faster than $A\propto N^2$ (although in this
case the $N>100$ region is numerically inaccessible).
This indicates that the QPT behavior of the specific heat (\ref{heat}) 
is consistent with the behavior of the standard specific heat in 
the first- and second-order thermodynamic phase transitions, in 
agreement with the above proposed conjecture \cite{Cej5}.

This conclusion is also supported by the analysis of the 
$N\to\infty$ limit.
Indeed, due to the degeneracy of $\beta_0=0$ and $\beta_0\neq 0$ 
minima of the function (\ref{Efun}) at $\lambda_{\rm c}$ for
$\chi\neq 0$, there will be a branch point on g.s. Riemann sheet 
that asymptotically reaches the real axis.
In contrast, there is no actual ground-state involving degeneracy 
in the $N\to\infty$ limit of the second-order transition, although, 
as can be shown, exceptional points come infinitely close to the 
real axis at the critical point.

Let us stress that in calculations leading to the above-discussed 
results all levels with $J=0$ were included in Eq.~(\ref{heat}).
We know, however, that for $\chi=0$ some pairs of levels actually 
cross at real $\lambda$ (due to the seniority quantum number), which 
implies that first and second derivatives of the corresponding
energies may be discontinuous or diverging.
Fortunately, a detailed consideration of this problem reveals that 
all these singularities cancel out exactly and do not affect the 
shapes of curves in Figs.~\ref{figheat}(b) and \ref{asymp}.
Indeed, if $E_i$ crosses with $E_{i+1}$, the discontinuity (singularity) 
of the first (second) derivatives in Eq.~(\ref{heat}) for both levels 
is the same, but with opposite signs, so the sum of both contributions 
is incremented correctly, as if the levels continuously passed the 
crossing.

In any case, one can repeat all $\chi=0$ calculations including only 
the $v=0$ subspace of states with $J=0$ into expression (\ref{heat}).
Since this subspace contains the ground state and does not mix with
$v\neq 0$ subspaces, one obtains a plausible new definition of the
$\chi=0$ specific heat, which avoids the above-discussed problem with 
singularities.
The dimension of the $v=0$ subspace grows as $n\sim N/2$, thus 
$\Omega\propto N$. 
Maximal values of the resulting specific heat curves with this
normalization are shown in Fig.~\ref{asymp} by triangles.
We observe a similar dependence as before, but because of lower
dimensions the convergence with $N$ is much slower now than before, 
not allowing a fully conclusive evidence of finite asymptotics.
The available data for boson numbers up to 5000 (see the inset in
Fig.~\ref{asymp}) are consistent with large-$N$ behaviors given 
by $A\propto N^k$ for $k\in(0,1.35)$, in agreement with 
the expected value $k=1$.
Let us stress that the dimension $n$ of the $\nu=0$ subspace at
$N=5000$ is comparable with the dimension of the whole $J=0$ 
subspace at $N\approx 170$, so the absence of saturation in the 
$\nu=0$ case in the available domain of boson numbers is not
surprising.
To obtain $\nu=0$ results comparable with $J=0$ at $N\approx 300$, 
one would need to go up to $N\approx 15000$, which is beyond our
present computational possibilities.

\section{Conclusions and outlook}

We have discussed several aspects of quantum structural phase 
transitions at zero temperature that could eventually lead to
their unified description with classical thermodynamic phase 
transitions.
The most fundamental aspect seems to follow from the analogy 
\cite{Cej5} between
the branch points related to the ground-state Riemann-sheet of
Hamiltonian (\ref{Hamg}) and complex zeros of the partition
function of a classical phase-transitional system.
Numerical data strongly supporting this conjecture were obtained in 
the interacting boson model, although a fully conclusive test would
require to further increase the upper limit of available boson numbers.
The immediate next task is to extend the calculations presented here
to other systems that exhibit quantum phase transitions of various 
orders.

An open problem is the generalization of our approach to finite
temperatures.
In this case, both structural and thermodynamic phase transitions 
describe nonanalytic behaviors of the thermodynamic potential 
$F_0(\lambda,T)$ along two perpendicular directions in the 
$\lambda\times T$ parameter space.
While the $T$-direction is described by standard thermodynamics,
involving the theory of complex zeros of the partition function,
the $\lambda$-direction is not as familiar.
Formula (\ref{finiT}) represents a possible starting point of 
the analysis. 
The finite-temperature average of $V'$ is just a weighted sum of averages 
corresponding to individual excited states
\begin{equation}
\ave{V'}{T}=\frac{1}{Z}\,\sum_i e^{-T^{-1}E_i}\,\ave{V'}{i}
\label{finiTe}
\end{equation}
and one can ask whether a thermally weighted sum $\rho_T$ of branch-point 
densities $\rho_i$ on the Riemann sheets assigned to excited states
will, in the finite-$T$ case, take the role of the g.s. density $\rho_0$, 
as discussed above. 

The interacting boson model seems to be an ideal tool for studying 
various aspects of complex quantum dynamics, including quantum 
phase transitions.
One of its advantages in this respect is the presence of quantum phase
transitions of both first and second orders.
We consider this model as one of the best candidates also for future
finite-temperature studies.

\section{Acknowledgments}
P.C. and S.H. thank Jan Jolie for relevant discussions.
This work was supported by GA{\v C}R and AS{\v C}R under
Project Nos. 202/02/0939 and K1048102, respectively, and
by the DFG under Grant No. 436 TSE 17/6/03.


\begin{thebibliography}{99}
\bibitem{Sachdev} S. Sachdev, {\it Quantum Phase Transitions\/} (Cambridge
 University Press, Cambridge, UK, 1999).
\bibitem{Gil} R. Gilmore, {\it Catastrophe Theory for Scientists and
 Engineers\/} (Wiley, New York, 1981).
\bibitem{Diep} A.E.L. Dieperink, O. Scholten, and F. Iachello, Phys.
 Rev. Lett. {\bf 44}, 1747 (1980).
\bibitem{Feng} D.H. Feng, R. Gilmore, and S.R. Deans, Phys. Rev. C
 {\bf 23}, 1254 (1981).
\bibitem{Zha} W.-M. Zhang, D.H. Feng, and J.N. Ginocchio, Phys. Rev.
 Lett. {\bf 59}, 2032 (1987).
\bibitem{Heiss1} W.D. Heiss, Z. Phys. A - Atomic Nuclei {\bf 329},
 133 (1988); W.D. Heiss and A.L. Sannino, Phys. Rev. A {\bf 43},
 4159 (1991); W.D. Heiss, Phys. Rep. {\bf 242}, 443 (1994).
\bibitem{Mor} E. L{\'o}pez-Moreno and O. Casta{\~n}os, Phys. Rev. C
 {\bf 54}, 2374 (1996).
\bibitem{Rowe} D.J. Rowe, C. Bahri, and W. Wijesundera, Phys. Rev. Lett.
 {\bf 80}, 4394 (1998).
\bibitem{Cas} R.F. Casten, D. Kusnezov, and N.V. Zamfir, Phys. Rev. Lett. 
 {\bf 82}, 5000 (1999).
\bibitem{Cej1} P. Cejnar and J. Jolie, Phys. Rev. E {\bf 61}, 6237 (2000).
\bibitem{Ia1} F. Iachello, Phys. Rev. Lett. {\bf 85}, 3580 (2000);
 {\it ibid.\/} {\bf 87}, 052502 (2001).
\bibitem{Cej2} P. Cejnar, V. Zelevinsky, and V.V. Sokolov, Phys. Rev. E
 {\bf 63}, 036127 (2001).
\bibitem{Jol1} J. Jolie, R.F. Casten, P. von Brentano, and V. Werner,
 Phys. Rev. Lett. {\bf 87}, 162501 (2001).
\bibitem{Jol2} J. Jolie, P. Cejnar, R.F. Casten, S. Heinze, A. Linnemann,
 and V. Werner, Phys. Rev. Lett. {\bf 89}, 182502 (2002). 
\bibitem{Cejle} P. Cejnar, Phys. Rev. Lett. {\bf 90}, 112501 (2003).
\bibitem{Cej4} P. Cejnar, S. Heinze, and J. Jolie, Phys. Rev. C {\bf 68}, 
 034326 (2003).
\bibitem{Vol} A. Volya and V. Zelevinsky, Phys. Lett. B {\bf 574}, 27 (2003).
\bibitem{Ema} C. Emary and T. Brandes, Phys. Rev. Lett. {\bf 90}, 044101 
 (2003); N. Lambert, C. Emary, T. Brandes, {\it ibid.} {\bf 92}, 073602 
 (2004).
\bibitem{Ari} J.M. Arias, J. Dukelsky, and J.E. Garc{\'\i}a-Ramos,
 Phys. Rev. Lett. {\bf 91}, 162502 (2003).
\bibitem{IaZ} F. Iachello and N.V. Zamfir, Phys. Rev. Lett. {\bf 92},
 212501 (2004).
\bibitem{Jol3} J. Jolie, S. Heinze, P. Van Isacker, and R.F. Casten,
 Phys. Rev. C {\bf 70}, 011305(R) (2004).
\bibitem{Rowe2} D.J. Rowe, Phys. Rev. Lett. {\bf 93}, 122502 (2004);
 D.J. Rowe, P.S. Turner, and G. Rosensteel, {\it ibid.\/} {\bf 93},
 232502 (2004).
\bibitem{Cap} J.M. Arias, J.E. Garc{\'\i}a-Ramos, and J. Dukelsky,
 Phys. Rev. Lett. {\bf 93}, 212501 (2004); M.A. Caprio and F. Iachello,
 {\it ibid.\/} {\bf 93}, 242502 (2004).
\bibitem{Cej5} P. Cejnar, S. Heinze, and J. Dobe{\v s}, Phys. Rev. C
 in press (2005); see also nucl-th/0406060.
%
\bibitem{Zel} V. Zelevinsky and A. Volya, Phys. Rep. {\bf 391}, 311 
 (2004).
%
\bibitem{Yang} C.N. Yang and T.D. Lee, Phys. Rev. {\bf 87}, 404 (1952);
 {\bf 87}, 410 (1952).
\bibitem{Gross} S. Grossmann and W. Rosenhauer, Z. Phys. {\bf 207}, 
 138 (1967); {\bf 218}, 437 (1969); S. Grossmann and V. Lehmann, 
 {\it ibid.} {\bf 218}, 449 (1969). 
%
\bibitem{Iach} F. Iachello, A. Arima, {\it The Interacting Boson
 Model\/} (Cambridge University Press, Cambridge, UK, 1987).
\bibitem{Gin} J.N. Ginocchio and M.V. Kirson, Nucl. Phys. {\bf A350},
 31 (1980).
%
\bibitem{Dys} F.J. Dyson, J. Math. Phys. {\bf 3}, 140 (1962).
\bibitem{Pec} P. Pechukas, Phys. Rev. Lett. {\bf 51}, 943 (1983);
 T. Yukawa, {\it ibid.} {\bf 54}, 1883 (1985).
%
\bibitem{Kato} T. Kato, {\it Perturbation Theory for Linear
 Operators\/} (Springer, Berlin, 1966).
\bibitem{Zirn} M.R. Zirnbauer, J.J.M. Verbaarschot, and H.A.
 Weidenm{\"u}ller, Nucl. Phys. {\bf A411}, 161 (1983).
\bibitem{Shan} P.E. Shanley, Ann. Phys. {\bf 186}, 292 (1988).
\bibitem{Heiss2} W.D. Heiss and W.-H. Steeb, J. Math. Phys. {\bf 32}, 
 3003 (1991). 
\bibitem{Rot2} I. Rotter, Phys. Rev. C {\bf 64}, 034301 (2001);
 I. Rotter and A.F. Sadreev, Phys. Rev. E {\bf 69}, 066201 (2004).
%
\bibitem{Levi} A. Leviatan, A. Novoselsky, and I. Talmi, Phys. Lett.
 B {\bf 172}, 144 (1986).
%
\bibitem{Bor} P. Borrmann, O. M{\"u}lken, and J. Harting, Phys. Rev.
 Lett. {\bf 84}, 3511 (2000); O. M{\"u}lken, H. Stamerjohanns, and
 P. Borrmann, Phys. Rev. E {\bf 64}, 047105 (2001).
\end{thebibliography}
\end{document}